\documentclass{article}
\usepackage{geometry}
\usepackage{amsmath}
\usepackage{amssymb}
\usepackage{multirow}

\begin{document}

\title{Quantum advice enhances social optimality in three-party conflicting interest games\thanks{The final publication is available at Rinton Press via Quantum Information \& Computation, Vol. 16, No. 7\&8 (2016) 0588-0596}}

\date{}

\author{Haozhen~Situ\thanks{Corresponding author: situhaozhen@gmail.com. College of Mathematics and Informatics, South China Agricultural University, Guangzhou 510642, China},
Cai~Zhang\thanks{College of Mathematics and Informatics, South China Agricultural University, Guangzhou 510642, China},
Fang~Yu\thanks{Department of Computer Science, Jinan University, Guangzhou 510632, China}}

\maketitle

\begin{abstract}
Quantum pseudo-telepathy games are good examples of explaining the strangeness of quantum mechanics
and demonstrating the advantage of quantum resources over classical resources.
Most of the quantum pseudo-telepathy games are common interest games,
nevertheless conflicting interest games are more widely used to model real world situations.
Recently Pappa \emph{et al.} (Phys. Rev. Lett. 114, 020401, 2015) proposed the first two-party conflicting interest game where quantum advice enhances social optimality.
In the present paper we give two new three-party conflicting interest games and show that quantum advice can enhance social optimality in a three-party setting.
The first game we propose is based on the famous GHZ game which is a common interest game.
The second game we propose is related to the Svetlichny inequality which demonstrates quantum mechanics cannot be explained by the local hidden variable model in a three-party setting.
\end{abstract}

\noindent\textbf{Keywords}: Quantum pseudo-telepathy game, nonlocal quantum game, multiparty game
\\\\

\section{Introduction}
Game theory is a useful analysis tool in many realms of scientific researches.
In the realm of quantum mechanics,
a lot of games (called quantum pseudo-telepathy games or nonlocal quantum games in the literatures, called quantum games for short in the present paper) have been proposed in order to illustrate the strangeness of quantum mechanics to non-physicists.
For example, Bell inequalities \cite{Bell} can be rephrased as games with incomplete information (also called Bayesian games) \cite{BL13}.
Usage of game theory terms like ``input, output, payoff, advice'' instead of physics terms like ``observable, measurement, eigenvalue, hidden variable, entanglement'' makes it easier for non-physicists to understand ``nonlocality'' \cite{nonlocality}, one of the strangest properties of quantum mechanics.
On the other hand, quantum games are good examples of demonstrating the advantage of quantum resources over classical resources
because quantum advice can enhance the players' payoffs in these games.
We remark that without more restrictions, quantum mechanics only offers advantages for Bayesian games, i.e., when the parties receive inputs or, in other words, when there are more than
a single measurement setting for each party \cite{Zhang}.

Most of the quantum games are common interest games
\cite{GHZ,magicsquare1,magicsquare2,hiddenmatching1,hiddenmatching2,CI08,Bridge,Iqbal15}.
Common interest means that even though each player's goal is to maximize his/her own payoff regardless of the other players', their interests have no conflict. Thus they tend to cooperate as a team when playing common interest games.
A team with quantum resources can outperform a team with only classical resources in quantum games.
But more often in the real world players have conflicting interests and
they tend to compete rather than cooperate.
Can quantum resources offer an advantage to conflicting interest games?
Recently, Pappa \emph{et al.} \cite{Pappa} proposed the first conflicting interest game where
quantum advice and quantum strategy can enhance social optimality.
They suggest that it would be interesting to find more conflicting interest games
where quantum mechanics offers an advantage, for example when larger dimensions are used or in a multiparty setting.
Later, Situ \cite{Situ} proposed a conflicting interest game with ternary outputs where
$3-$dimensional quantum systems are used to enhance social optimality.
A conflicting interest quantum game with ternary inputs is also given \cite{Situ}.

In the present paper, we try to find conflicting interest games where quantum mechanics can enhance social optimality in a three-party setting.
Specifically, we construct a new three-party conflicting interest game on the basis of the common interest GHZ game \cite{GHZ}, and show that quantum advice can enhance social optimality in this new game.
In the GHZ game, the combination of the three players' inputs are promised to belong to a subset of the product of the individuals' input sets.
It would be interesting to ask whether a similar game without this restriction still benefits from quantum advice.
The answer is affirmative.
We construct another three-party conflicting interest game without the input restriction and show that
quantum advice can also enhance the social optimality in this game.

The remainder of the present paper is organized as follows:
Section \ref{preliminaries} reviews the concepts of Bayesian game, advice and equilibirum.
Section \ref{GHZgame} reviews the original common interest GHZ game.
Section \ref{withpromise} gives our first game which is constructed on the basis of the GHZ game.
Section \ref{withoutpromise} gives our second game which is free from input restriction.
Section \ref{conclusion} is a brief conclusion.

\section{Preliminaries}\label{preliminaries}

Bayesian game is used to model situations in which some of the players are not certain of the characteristics of some of the other players.
Here we define a Bayesian game in the three-party framework (for a more general definition refer to Ref. \cite{gametheory}).
A three-party Bayesian game consists of:
\begin{itemize}
\item[-]
three players, here we call them Alice(A), Bob(B) and Charlie(C),
\end{itemize}
and for each player $i\in\{A,B,C\}$
\begin{itemize}
\item[-]
a set $\mathcal{X}_i$ of inputs $x_i$,
\item[-]
a set $\mathcal{Y}_i$ of outputs $y_i$,
\item[-]
a payoff function $\$_i: \mathcal{X}\times\mathcal{Y}\rightarrow \mathbb{R}$, where $\mathcal{X}=\mathcal{X}_A\times\mathcal{X}_B\times\mathcal{X}_C, \mathcal{Y}=\mathcal{Y}_A\times\mathcal{Y}_B\times\mathcal{Y}_C$, $\mathbb{R}$ denotes the set of real numbers
\end{itemize}
and
\begin{itemize}
\item[-]
a probability distribution $P: \mathcal{X}\rightarrow [0,1]$.
\end{itemize}
The games studied in the present paper all involve binary inputs and binary outputs, i.e., $\mathcal{X}_i=\mathcal{Y}_i=\{0,1\}$.
We introduce the notation $x=(x_A,x_B,x_C)$ and $y=(y_A,y_B,y_C)$ for convenience.
The game is played as follows. Each player $i\in\{A,B,C\}$ receives an input $x_i$,
and must produce an output $y_i$, respectively.
Each player only knows his/her own input but not the other players',
and they are not allowed to communicate with each other.
Each player's goal is to maximize his/her average payoff
\begin{align}
F_i=\sum_{x\in\mathcal{X}} P(x) \sum_{y\in\mathcal{Y}} Pr(y|x)\$_i(x,y),
\end{align}
where $Pr(y|x)$ is the probability of the players outputting $y$ given inputs $x$.

To play the game, each player should decide on a particular strategy.
A pure strategy for player $i$ consists in associating an output $y_i$ for every possible input $x_i$, i.e., a mapping $\mathcal{X}_i\rightarrow \mathcal{Y}_i$.
More generally, each player $i$ may use a mixed strategy, defined by $Pr(y_i|x_i)$,
which is the probability of outputting $y_i$ given the input $x_i$.
A pure strategy is also a mixed strategy satisfying that
for any $x_i\in\mathcal{X}_i$, there exists only one $y_i\in\mathcal{Y}_i$ such that $Pr(y_i|x_i)$ is nonzero.
If all players adopt mixed strategies the probability $Pr(y|x)$ is factorizable:
\begin{align}
Pr(y|x)=Pr(y_A|x_A)Pr(y_B|x_B)Pr(y_C|x_C).
\end{align}
A triple of strategies $(Pr(y_A|x_A),Pr(y_B|x_B),Pr(y_C|x_C))$ is an equilibrium \cite{gametheory} if
each player cannot obtain a higher average payoff by deviating from his/her strategy,
as long as the other players keep their strategies unchanged.

We also consider the case where the players receive advice from a source that is independent of the inputs $x$, and that can be classical or quantum.
This opens the possibility for the players to adopt correlated strategies,
which can outperform independent strategies.

In the case of classical advice, we define a correlated strategy $s_i: \mathcal{X}_i\times\Omega_i\rightarrow\mathcal{Y}_i$,
where $\Omega_i$ is the space of advice given to player $i$ by the source
and the source chooses the advice $r=(r_A,r_B,r_C)$ from the space $\Omega=\Omega_A\times\Omega_B\times\Omega_C$
following a probability distribution $\mathcal{Q}$.
Given an input $x_i$ and advice $r_i$, player $i$ chooses the output $y_i=s_i(x_i,r_i)$.
If all players follow a correlated strategy $s=(s_A,s_B,s_C)$,
then
\begin{align}
Pr(y|x) & =\sum_{r\in\Omega}\mathcal{Q}(r)Pr(y_A|x_A,r_A)Pr(y_B|x_B,r_B)Pr(y_C|x_C,r_C)\nonumber\\
& =\sum_{r\in\Omega}\mathcal{Q}(r)Pr\big(y_A\ \mathrm{equals}\ s_A(x_A,r_A)\big)Pr\big(y_B\ \mathrm{equals}\ s_B(x_B,r_B)\big)Pr\big(y_C\ \mathrm{equals}\ s_C(x_C,r_C)\big),
\end{align}
which is generally not factorizable.
A correlated strategy $s$ is a correlated equilibrium \cite{CE}
if player $i$ cannot gain a higher average payoff by changing his/her strategy unilaterally for any $i\in\{A,B,C\}$.

In the case of quantum advice, we define a quantum strategy $\mathcal{M}=(\mathcal{A},\mathcal{B},\mathcal{C},|\Psi\rangle)$,
with $\mathcal{A}=(\mathcal{A}_0,\mathcal{A}_1)$,
$\mathcal{B}=(\mathcal{B}_0,\mathcal{B}_1)$ and
$\mathcal{C}=(\mathcal{C}_0,\mathcal{C}_1)$.
The players apply, respectively, the observables $\mathcal{A}_{x_A}=\{A_{x_A}^0,A_{x_A}^1\}$,
$\mathcal{B}_{x_B}=\{B_{x_B}^0,B_{x_B}^1\}$ and $\mathcal{C}_{x_C}=\{C_{x_C}^0,C_{x_C}^1\}$
on the shared quantum state or advice $|\Psi\rangle$.
The probability of the players outputting $y$ given $x$ is
\begin{align}
Pr(y|x)=\langle\Psi| A_{x_A}^{y_A}\otimes B_{x_B}^{y_B}\otimes C_{x_C}^{y_C} |\Psi\rangle,
\end{align}
which can in general not be reproduced by any classical local model.
A quantum strategy $\mathcal{M}$ is a quantum equilibrium \cite{Pappa} if no player can gain a higher average payoff by choosing a different strategy unilaterally.

\section{The GHZ game}\label{GHZgame}

The famous GHZ game \cite{GHZ} is widely used to illustrate the strangeness of quantum mechanics in many literatures.
It involves three players, Alice, Bob and Charlie.
Each player $i\in\{A,B,C\}$ has the input set $\mathcal{X}_i=\{0,1\}$, and $\mathcal{X}=\mathcal{X}_A\times\mathcal{X}_B\times\mathcal{X}_C$ follows the probability distribution $P(0,0,0)=P(0,1,1)=P(1,0,1)=P(1,1,0)=1/4$.
We can see that the inputs satisfy condition $x_A\oplus x_B \oplus x_C=0$.
The output set for each player $i$ is $\mathcal{Y}_i=\{0,1\}$.
The payoff function $\$_i$ for player $i$ is defined as follows:
\begin{align}
\$_i(0,0,0,y_A,y_B,y_C)=\begin{cases}
1,    & y_A\oplus y_B \oplus y_C=0  \\
0,    & y_A\oplus y_B \oplus y_C=1
\end{cases},\\
\$_i(x,y_A,y_B,y_C)=\begin{cases}
1,    & y_A\oplus y_B \oplus y_C=1  \\
0,    & y_A\oplus y_B \oplus y_C=0
\end{cases},
\end{align}
for every $x\in\{011,101,110\}$, where $\oplus$ denotes addition modulo 2.
Because the payoff functions for all players are the same, they have common interest.

It's well known that there's no way for the three players to obtain a payoff of 1 for sure by classical advice. However, if they are able to share a GHZ state,
they can choose suitable measurement settings according to the inputs and then output their measurement outcomes of their particles,
which always result in a payoff of 1.
Having access to quantum advice, the team can obtain a payoff of 1 with certainty.
Therefore the GHZ game demonstrates that shared entanglement can generate some correlations that cannot be explained by the local hidden variable model in a three-party setting.

\section{Three-party conflicting interest quantum game with promised inputs}\label{withpromise}

This section presents our first three-party conflicting interest quantum game where the inputs satisfy $x_A\oplus x_B \oplus x_C=0$, like the GHZ game.
Our game has the same structure as the GHZ game,
except for different payoff functions $\$_A,\$_B,\$_C$ for Alice, Bob and Charlie.
We define the three payoff functions as follows:
\begin{align}
\label{eq_GHZ_payoff1}
& \$_{A,B,C}(0,0,0,y_A,y_B,y_C)=\begin{cases}
(\frac{4}{3},\frac{4}{3},\frac{4}{3}),    & y_A\oplus y_B\oplus y_C=0  \\
0,    & y_A\oplus y_B\oplus y_C=1
\end{cases},\\
\label{eq_GHZ_payoff2}
& \$_{A,B,C}(x,y_A,y_B,y_C)=\begin{cases}
(2,1,1), & (y_A,y_B,y_C)=(1,0,0)\\
(1,2,1), & (y_A,y_B,y_C)=(0,1,0)\\
(1,1,2), & (y_A,y_B,y_C)=(0,0,1)\\
(\frac{4}{3},\frac{4}{3},\frac{4}{3}), &  (y_A,y_B,y_C)=(1,1,1)\\
(0,0,0), & y_A\oplus y_B\oplus y_C=0
\end{cases},
\end{align}
for every $x\in\{011,101,110\}$.
The values in the triples refer to Alice, Bob and Charlie's payoffs respectively.

In the absence of advice,
equilibria with average payoffs $(\frac{13}{12},\frac{5}{6},\frac{13}{12})$ and $(\frac{13}{12},\frac{13}{12},\frac{5}{6})$ fulfil Alice's preference,
 equilibria with average payoffs $(\frac{5}{6},\frac{13}{12},\frac{13}{12})$ and $(\frac{13}{12},\frac{13}{12},\frac{5}{6})$ fulfil Bob's preference,
 equilibria with average payoffs $(\frac{5}{6},\frac{13}{12},\frac{13}{12})$ and $(\frac{13}{12},\frac{5}{6},\frac{13}{12})$ fulfil Charlie's preference.
However, no one  equilibrium can fulfil all three players' preferences,
so conflicting interests rise.
In the case of classical advice, the set of all possible triples of average payoffs $(F_A,F_B,F_C)$
forms a convex polytope in $\mathcal{R}^3$.
We can therefore examine all possible strategies and see that
$F_A+F_B+F_C\leq 3$,
which is of course satisfied by all possible correlated equilibria.
The social optimal outcome is at most 3.

Now we add up all the three players' payoffs to get the total payoff
\begin{align}
& \$_{total}(0,0,0,y_A,y_B,y_C)=\begin{cases}
4,    & y_A\oplus y_B\oplus y_C=0  \\
0,    & y_A\oplus y_B\oplus y_C=1
\end{cases},\\
& \$_{total}(x,y_A,y_B,y_C)=\begin{cases}
4, & y_A\oplus y_B\oplus y_C=1\\
0, & y_A\oplus y_B\oplus y_C=0
\end{cases},
\end{align}
for every $x\in\{011,101,110\}$.
This total payoff function is the same as the common payoff function of the original GHZ game except for a factor of $4$.
If the players are able to share a GHZ state $|\Psi\rangle=\frac{1}{\sqrt{2}}(|000\rangle+|111\rangle)$,
they can measure their particles in a suitable measurement setting according to the inputs,
and then output the measurement outcomes.
Because the measurement outcomes always result in a payoff of 4 no matter what the inputs are,
using quantum advice enhances the total average payoff from $3$ to $4$,
a better social optimal outcome is achieved.

For completeness we include the measurement settings \cite{GHZ} used in this game.
If a player receives an input $0$, he/she will measure his/her particle in the $|+\rangle=\frac{1}{\sqrt{2}}(|0\rangle+|1\rangle),|-\rangle=\frac{1}{\sqrt{2}}(|0\rangle-|1\rangle)$
basis (the eigenstates of the Pauli $\sigma_X$ operator) and if he/she gets outcome $|+\rangle$, then he/she outputs $0$ and if he/she gets outcome $|-\rangle$, then he/she outputs $1$.
Similarly if a player receives an input $1$, he/she measures his/her particle in the
$|+i\rangle=\frac{1}{\sqrt{2}}(|0\rangle+i|1\rangle),|-i\rangle=\frac{1}{\sqrt{2}}(|0\rangle-i|1\rangle)$
basis (the eigenstates of the Pauli $\sigma_Y$ operator) and if he/she gets outcome
$|+i\rangle$, then he/she outputs $0$ and if he/she gets outcome $|-i\rangle$,
then he/she outputs $1$.
By straightforward calculation we have
$Pr(1,0,0|x)=Pr(0,1,0|x)=Pr(0,0,1|x)=Pr(1,1,1|x)=1/4$
for every $x\in\{011,101,110\}$, thus the average payoffs for each player
is $\frac{1}{4}(2+1+1+\frac{4}{3})=\frac{4}{3}$. The solution is fair.

A further question is whether the above quantum fair strategies constitute an equilibrium?
The answer is affirmative. We first prove that Alice cannot obtain an average payoff higher than $\frac{4}{3}$ by unilaterally deviating from her strategy as long as Bob and Charlie keep their strategies unchanged.
We distinguish four cases according to the inputs $x$.
In the case $x=(0,0,0)$, Alice's maximal payoff defined in Eq. (\ref{eq_GHZ_payoff1}) is $\frac{4}{3}$. No matter what measurements the three players perform, Alice cannot obtain an average payoff higher than $\frac{4}{3}$.
In the case $x=(0,1,1)$, Bob and Charlie will both measure observable $\sigma_Y$.
By straightforward calculation we can see that
Bob and Charlie will get measurement outcomes $(0,0)$,$(1,0)$,$(0,1)$,$(1,1)$ with equal probability.
By referring to Eq. (\ref{eq_GHZ_payoff2}), we can see that the upper bound of Alice's average payoff is
$\frac{1}{4}(2+1+1+\frac{4}{3})=\frac{4}{3}$.
The distribution of Bob and Charlie's measurement outcomes are independent of Alice's choice of measurement because of the no-signalling condition,
so Alice cannot obtain an average payoff higher than $\frac{4}{3}$ by unilateral deviation.
In the other two cases $x=(1,0,1)$ and $x=(1,1,0)$, same conclusions can be made.
So we complete the proof that Alice cannot obtain an average payoff higher than $\frac{4}{3}$ by unilateral deviation.
In the same way we can prove that neither Bob nor Charlie can obtain an average payoff higher than $\frac{4}{3}$ by unilateral deviation.
So we complete the proof that the above quantum fair strategies constitute an equilibrium.

\section{Three-party conflicting interest quantum game with all inputs}\label{withoutpromise}

In the previous section we construct a conflicting interest quantum game on the basis of the common interest GHZ game where the inputs are promised to belong to a subset $\{000,011,101,110\}$ of $\mathcal{X}=\{0,1\}\times\{0,1\}\times\{0,1\}$. Now we come to the question whether quantum advice can also enhance social optimality in a similar three-party conflicting interest game where all inputs $x\in\mathcal{X}$ are allowable.
The answer is affirmative and we give an example in this section.
Let $P(x_A,x_B,x_C)=1/8$ for every $(x_A,x_B,x_C)\in \mathcal{X}_A\times\mathcal{X}_B\times\mathcal{X}_C$,
and define the payoff functions as follows:
\begin{align}
& \$_{A,B,C}(0,0,0,y_A,y_B,y_C)=\begin{cases}
(\frac{4}{3},\frac{4}{3},\frac{4}{3}),    & y_A\oplus y_B \oplus y_C=0   \\
0,    & y_A\oplus y_B \oplus y_C=1
\end{cases},\\
& \$_{A,B,C}(1,1,1,y_A,y_B,y_C)=\begin{cases}
(\frac{4}{3},\frac{4}{3},\frac{4}{3}),    & y_A\oplus y_B \oplus y_C=1  \\
0,    & y_A\oplus y_B \oplus y_C=0
\end{cases},\\
& \$_{A,B,C}(x,y_A,y_B,y_C)=\begin{cases}
(2,1,1), & (y_A,y_B,y_C)=(1,0,0)\\
(1,2,1), & (y_A,y_B,y_C)=(0,1,0)\\
(1,1,2), & (y_A,y_B,y_C)=(0,0,1)\\
(\frac{4}{3},\frac{4}{3},\frac{4}{3}), &  (y_A,y_B,y_C)=(1,1,1)\\
(0,0,0), & y_A\oplus y_B \oplus y_C=0
\end{cases},\\
& \$_{A,B,C}(x',y_A,y_B,y_C)=\begin{cases}
(2,1,1), & (y_A,y_B,y_C)=(0,1,1)\\
(1,2,1), & (y_A,y_B,y_C)=(1,0,1)\\
(1,1,2), & (y_A,y_B,y_C)=(1,1,0)\\
(\frac{4}{3},\frac{4}{3},\frac{4}{3}), &  (y_A,y_B,y_C)=(0,0,0)\\
(0,0,0), & y_A\oplus y_B \oplus y_C=1
\end{cases},
\end{align}
for every $x\in\{011,101,110\}$ and $x'\in\{100,010,001\}$.

In the absence of advice,
equilibria with average payoffs $(\frac{7}{6},\frac{11}{12},\frac{11}{12})$ fulfil Alice's preference,
equilibria with average payoffs $(\frac{11}{12},\frac{7}{6},\frac{11}{12})$ fulfil Bob's preference,
equilibria with average payoffs $(\frac{11}{12},\frac{11}{12},\frac{7}{6})$ fulfil Charlie's preference.
However, no one equilibrium can fulfil all the three  players' preferences, so they have conflicting interests.
In the case of classical advice,
we have $F_A+F_B+F_C\leq 3$, which is of course satisfied by all possible correlated equilibria.
The social optimal outcome is at most 3.

We add up all the three players' payoffs to get the total payoff
\begin{align}
& \$_{total}(x,y_A,y_B,y_C)=\begin{cases}
4, & y_A\oplus y_B \oplus y_C=1 \\
0, & y_A\oplus y_B \oplus y_C=0
\end{cases},\\
& \$_{total}(x',y_A,y_B,y_C)=\begin{cases}
4, & y_A\oplus y_B \oplus y_C=0 \\
0, & y_A\oplus y_B \oplus y_C=1
\end{cases},
\end{align}
for every $x\in K_1=\{111,011,101,110\}$ and $x'\in K_0=\{000,100,010,001\}$.
The total average payoff function can be written as
\begin{align}
F_{total} & =\sum_{x\in K_0\cup K_1}P(x)\sum_{y\in\mathcal{Y}} Pr(y|x)\$_{total}(x,y)
\nonumber\\  =& \frac{1}{8}\sum_{x\in K_0}\sum_{y_A\oplus y_B \oplus y_C=0} Pr(y|x)\cdot 4 + \frac{1}{8}\sum_{x\in K_1}\sum_{y_A\oplus y_B \oplus y_C=1} Pr(y|x)\cdot 4
\nonumber\\ =& \frac{1}{2}\sum_{x\in K_0}\frac{1+E(x)}{2} + \frac{1}{2}\sum_{x\in K_1}\frac{1-E(x)}{2}
\nonumber\\ =& 2+\frac{1}{4}\Big(\sum_{x\in K_0}E(x)-\sum_{x\in K_1}E(x)\Big)
\nonumber\\ =& 2+\frac{1}{4}\Big(E(000)+E(100)+E(010)+E(001)    -E(111)-E(011)-E(101)-E(110)\Big),
\end{align}
where we have defined
\begin{align}
E(x)= &\sum_{y_A\oplus y_B \oplus y_C=0}Pr(y|x)-\sum_{y_A\oplus y_B \oplus y_C=1}Pr(y|x)
\nonumber\\ = &2\sum_{y_A\oplus y_B \oplus y_C=0}Pr(y|x)-1
\nonumber\\ = &1-2\sum_{y_A\oplus y_B \oplus y_C=1}Pr(y|x).
\end{align}
$E(000)+E(100)+E(010)+E(001)-E(111)-E(011)-E(101)-E(110)$ is actually the Bell expression of the Svetlichny inequality \cite{Svetlichny}.
The classical bound for the Svetlichny inequality is $4$
and the quantum violation is approximately $5.656$.
If the players have access to quantum advice,
the total average payoff can be enhanced from $3$ to approximately $2+\frac{1}{4}\cdot 5.656\approx3.414$,
a better social optimal outcome is achieved.
We have checked that the quantum state and the measurement settings \cite{Svetlichny} for the quantum violation
result in a fair allocation of payoff. The average payoffs for the three players
are approximately $(1.138,1.138,1.138)$.

Moreover, the above quantum fair strategies constitute an equilibrium.
This can be proved via semidefinite programming (SDP).
We first show that Alice cannot obtain a higher average payoff by unilaterally deviating from her strategy as long as Bob and Charlie keep their strategies unchanged.
In order to apply the technique in Ref. \cite{NPA07}, which employs SDP to bound bipartite quantum correlations,
we have to regard Bob and Charlie as one party that have the input set $\{00,01,10,11\}$ and the output set $\{00,01,10,11\}$.
Because Alice has two measurement choices that each yields one out of two outcomes,
there are $2\times2=4$ projection operators on Alice's side.
Bob and Charlie have four measurement choices that each yields one out of four outcomes,
so there are $4\times4=16$ projection operators on Bob and Charlie's side.
We put these 20 operators in a set $\mathcal{S}=\{S_1,\ldots,S_{20}\}$.
Then a $20\times20$ matrix $\Gamma$ is associated to the set $\mathcal{S}$ through $\Gamma_{ij}=\mathrm{Tr}(S_i^\dagger S_j\rho)$.
We fix the state $\rho$ and Bob and Charlie's measurements and fill in the matrix $\Gamma$ with determined values
$\mathrm{Tr}(S_i^\dagger S_j\rho)$ with $S_i, S_j$ belonging to Bob and Charlie's side.
We also impose the constraints that probabilities are nonnegative and sum to 1, as well as the no-signalling condition.
Finally we maximize Alice's average payoff subject to $\Gamma\geq0$.
The resulted upper bound is approximately $1.138$.
So we can infer that Alice cannot obtain a higher average payoff by unilateral deviation.
In the same way we can check that neither Bob nor Charlie can obtain a higher average payoff by unilateral deviation.
So the quantum fair strategies constitute an equilibrium.

\section{Conclusion}\label{conclusion}

Games have been widely used to explain quantum mechanics and demonstrate the advantage of quantum mechanics.
Conflicting interest games are seldom studied in the context of quantum pseudo-telepathy games.
Recently some authors \cite{Pappa,Situ} have given new examples of two-party conflicting interest games where quantum advice enhances social optimality.
The present paper has given two new three-party conflicting interest games of this genre.
The first one is related to the original common interest GHZ game \cite{GHZ},
while the second one is related to the Svetlichny inequality \cite{Svetlichny}.
The future work is to find more conflicting games where quantum mechanics offers an advantage, for example when larger than 3-dimensional quantum systems are used or
in a more than 3-party setting.
\\\\
\noindent\textbf{Acknowledgement}\\
This work is supported by the National Natural Science Foundation of China (Grant Nos. 61502179, 61472452) and the Natural Science Foundation of Guangdong Province of China (Grant Nos. 2014A030310265).

\end{document}